\documentclass[11pt,a4paper]{article}
\usepackage{ifpdf}
\usepackage{ae}
\usepackage[T1]{fontenc}
\usepackage[ansinew]{inputenc}
\usepackage{mathrsfs}
\usepackage{amsmath}
\usepackage{amssymb}
\pdfoutput=1
 \usepackage{
a4wide,graphicx,bm,times,psfrag,wrapfig,sidecap} \usepackage{cite}
\usepackage[colorlinks=true,linkcolor=black, citecolor=black,
urlcolor=black]{hyperref} \numberwithin{equation}{section}
\makeatletter \let\old@startsection=\@startsection
\renewcommand{\@startsection}[6]
{\old@startsection{#1}{#2}{#3}{#4}{#5}{#6\mathversion{bold}}} \def\Zh{
Zhukovsky\ } \makeatother \def\O{\Omega} \def\Det{ \text{Det}} 
\def\defeq{\stackrel{\text{def}}=} \newcommand\re[1]{({\ref{#1}})}
\def\be{\begin{eqnarray} } \def\ee{\end{eqnarray}} 
\def\IR{{\mathbb{R}}}  \def\IK{{\mathbb{K}}}
  \def\IN{{\mathbb{N}}}
 \def\ID{{\mathbb{D}}}
 \def\psu{{\mathfrak{psu} }}

\def\xp{{x^+ }}\def\xm{{x^-}}
\def\yp{{y^+}}\def\ym{{y^-}}

    \def\no{\nonumber}
    \def\la{\label}
    \def\l {\lambda}
\def\({\left(} \def\){\right)} 
\def\<{\langle\,} 
\def\>{\,  \rangle} 
\def\[{\left[}
 \def\]{\right]} 
\def\tr{{\rm   tr} }

     \def\CS{{\cal S}}

    \def\CZ{{ \mathcal{ Z} }}
    \def\CK{{ \mathcal{ K} }}
     
    \def\CI{{ \mathcal{ I} }}
     \def\CX{{ \mathcal{ X} }}
   \def\o{\omega} \def\CC{ {\mathcal C}}   \def\CN{{ \cal N}}    
     \def\a{\alpha}
  \def\b{\beta} \def\g{\gamma} \def\e{\epsilon} 
   \def\t{\tau}  \def\th{\theta}
  \def\G{\Gamma} 
  
 \def\Tr{{\rm Tr}}
\def\d{\delta}

   \def\sosix{{\mathfrak{so}(6)}} 
\newcommand\encadremath[1]{\vbox{\hrule\hbox{\vrule\kern8pt
\vbox{\kern8pt \hbox{$\displaystyle #1$}\kern8pt}
\kern8pt\vrule}\hrule}} \def\enca#1{\vbox{\hrule\hbox{
\vrule\kern8pt\vbox{\kern8pt \hbox{$\displaystyle #1$} \kern8pt}
\kern8pt\vrule}\hrule}}

  \usepackage{bm}
\def\O{\Omega} \def\defeq{\stackrel{\text{def}}=}
\def\ee{\end{eqnarray}}  \def\IR{{\mathbb{R}}}
  \def\IN{{\mathbb{N}}}
    \def\no{\nonumber} \def\la{\label} \def\l {\lambda}
\def\({\left(} \def\){\right)}    \def\[{\left[} \def\]{\right]} \def\tr{{\rm tr} }
 
 \def\CS{{\cal S}}    \def\CZ{{ \mathcal{ Z} }}
\def\CK{{ \mathcal{ K} }} \def\CI{{ \mathcal{ I} }}

\def\Tr{{\rm Tr}}  \def\d{\delta}

     \def\CP{{\cal P}}

 \def\CKK{\bm{\CK}} \def\CCC{\bm{\CC}}
 
\def\Psib{\bm{\Psi}} 
\def\CR{\mathcal{R}} \def\CRR{ \bm{\mathcal{R}}} \def\KK{ \bm{K}}
\def\JJ{ \bm{J}} \def\II{ \bm{I}} \def\pf{\mathrm{Pf}}
  \def\LL{
\mathbf{L}}\def\RR{ \mathbf{R}} \def\C{\mathbf{C}}
\def\K{\mathrm{K}} \def\KKK{\mathbf{K}} \def\Phib{
{\bf \Phi}} 

\begin{document}
\vspace*{-.4in} \thispagestyle{empty} \vspace{.1in} {\Large
\begin{center}
{\bf 
The Octagon as a Determinant
}
\end{center}}

\begin{center}
\vspace{8mm}

Ivan Kostov$^a$, Valentina B. Petkova$^b$, Didina Serban$^a$

  \vskip 6mm

\small{

{\textit{ $^a$ Institut de Physique Th\'eorique, DSM, CEA, URA2306
CNRS\\Saclay, F-91191 Gif-sur-Yvette, France}}

 {\it $^b$ Institute for Nuclear Research and Nuclear Energy,
 \\
Bulgarian Academy of Sciences, Sofia, Bulgaria}


}

\bigskip

 {\small\tt ivan.kostov@ipht.fr,\ petkova@inrne.bas.bg,
 didina.serban@ipht.fr }

\end{center}

\begin{abstract}
\normalsize{ The computation of a certain class of four-point
functions of heavily charged BPS operators boils down to the
computation of a special form factor - the octagon.  In this paper,
which is an extended version of the short note \cite{Kostov:2019stn},
we derive a non-perturbative formula for the square of the octagon as
the determinant of a semi-infinite skew-symmetric matrix.  We show
that perturbatively in the weak coupling limit the octagon is given by
a determinant constructed from the polylogarithms evaluating ladder
Feynman graphs.  We also give a simple operator representation of the
octagon in terms of a vacuum expectation value of massless free bosons
or fermions living in the rapidity plane.  }
\end{abstract}


\newpage

\setcounter{page}{1}
\begingroup
\hypersetup{linkcolor=black}
\endgroup

\section{Introduction}
\label{sec:intro}

In this paper we address the computation of the octagon form factor
which appears as a building block for the evaluation of a class of
four-point correlation functions of single-trace half-BPS operators in
$\CN=4$ planar SYM \cite{Coronado:2018ypq}.  We are doing so using
integrability-inspired techniques
\cite{Minahan:2002ve,Integrability-overview-2012,Gromov:2013pga,
Gromov:2014caa}, and in particular the recently proposed geometric
decomposition of correlation functions into hexagons
\cite{BKV1,Fleury:2016ykk,Eden:2016xvg, Fleury:2017eph,
Bargheer:2017nne, Bargheer:2018jvq}.  The hexagons are form factors
for non-local operators creating curvature defects.  While the
three-point correlation functions of half-BPS operators are trivial,
the four-point function of such operators have a rather rich
structure.

As pointed out in \cite{Coronado:2018ypq}, in certain cases these
correlation functions can be evaluated exactly.  In particular this is
the case in the limit of half-BPS operators with large R-charges and
specially tuned polarisations the correlation function factorises into
a sum of products of two octagon form factors, or octagons.  From the
point of view of the dual string theory, the octagon is an off-shell
open string partition function with classical boundaries representing
geodesics in AdS$_5$.  The closed string worldsheet representing a
sphere with four punctures contains four consecutive classical
geodesics.  Such a worldsheet splits into two open-string worldsheets
spanned on the four geodesics.  Very recently it was discovered that
similar factorisation occurs in any given order of the $1/N_c$
expansion \cite{Bargheer:2019kxb}.

 In this paper we obtain a closed analytic expression of the octagon
 at any coupling in terms of a Fredholm pfaffian.\footnote{The fact
 that the sum over virtual particles can be written as a Fredholm
 pfaffian has been noticed before in \cite{Basso:2017khq}.} This
 allows us to represent the square of the octagon as a Fredholm
 determinant.  To compute this Fredholm determinant, we write the
 Fredholm kernel $K(u,v)$ on a basis of functions $\psi_n(u)$,
 \be \la{degenerate} K(u,v) = \sum_{m, n=0}^\infty \psi_m(u)\ T_{mn} \
 \psi_n(v).  \ee with certain matrix operator $T$ and further
 factorise by Fourier transform.  If we restrict the sum to a finite
 ``cutoff'' $N$, the Fredholm deterrminant becomes a usual $N\times N$
 determinant.  It can happen, and this is the case in our problem,
 that the Fredholm determinant is given by the limit $N\to\infty$ of
 the ordinary determinant.  This procedure is similar to the method of
 degenerate kernels \cite{KaplanKaKry-book} used to solve integral
 equations.  It is also possible to approximate the Fredholm pfaffian
 by ordinary pfaffians by applying the pfaffian integration theorem
 \cite{Akemann:2007wa,1751-8121-40-36-F01}.  The advantage of this
 manipulation is that the multiple integrals break into sums of
 products of simple integrals.

 We claim that to any loop order the pfaffian can be recast as a
 determinant of another matrix.  From the determinant representations
 one confirms by direct computation the conjecture of
 \cite{Coronado:2018cxj} that the perturbative expansion of the
 octagon can be recast as a multilinear combination of ladder
 integrals
  \be
 \begin{aligned}
 \la{defInOa} 
 \mathbb{O} _{\ell}& =1+\sum_{n=1}^{\infty} \ \CX_n(\phi, \varphi,
 \th) \ \sum_{J=n(n+\ell)} ^\infty \ \sum_{ j_1+...j_n=J}
 c_{j_1,\cdots j_n} \ f_{j_1} \cdots f_{j_n } \ g^{2J }.
 \end{aligned}
 \ee
 In this sum the dependence on the polarisations is carried by the
 weight factors $\CX_n$ \cite{Fleury:2016ykk,Fleury:2017eph}, to be
 recalled in the following, the coefficients $c_{j_1...j_n}$ are
 rational numbers, and the (conveniently normalised) ladder integrals
 are given, for $|z|<1$, by
\be \la{deffk}
\begin{aligned}
f_k(z, \bar z) &\equiv k!(k-1)!  F_k(z,\bar z) \\
& = \sum_{j=k}^{2k} \frac{(k-1)!  \ j!  }{ (j-k)!  (2k-j)!} \ (-\log
z\bar z)^{2k-j} \ {\mathrm{Li}_{j}(z) - \mathrm{Li}_{j}(\bar z)\over
z-\bar z} .
\end{aligned}
\ee
The form \re{defInOa} of the perturbative octagon carries some
resemblance to the result of Basso and Dixon \cite{Basso:2017jwq}
obtained for the fishnet limit of the $\CN=4 $ SYM
\cite{Gurdogan:2015csr}\footnote{The integrability of the fishnet
Feynman graphs has been first established by A. Zamolodchikov
\cite{Zamolodchikov:1980mb}.}.  The analytic expression obtained in
\cite{Basso:2017jwq} for the fishnet has the form of a single
determinant of ladders, while bootstraping the Ansatz \re{defInOa} one
obtains \cite{Coronado:2018cxj} a series in the minors of the
semi-infinite matrix
  \be
  \mathbf{f}= 
  \left(
\begin{array}{cccccc}
 f_1 & f_2 & f_3 &  . \\
 f_2 & f_3 & f_4 & .  \\
 f_3 & f_4 & f_5  & . \\
. & . & . &.     \\
\end{array}
\right) .
\la{detf}
\ee

 Our explicit solution for the perturbative octagon is in the form of
 either pfaffian or determinant, of semi-infinite matrices with
 elements represented by series of ladder integrals.  This gives in
 particular an explicit analytic expression for the coefficients
 $c_{j_1,\cdots j_n} $ in the expansion
\re{defInOa}.  Finally, we give an alternative derivation of the
pfaffian and the determinant representations based on an operator
construction of the octagon with free real massless Majorana fermions
living on the \Zh plane.\footnote{By \Zh plane we mean the rapidity
plane with two simple branch points at $u=\pm 2g$.  A similar operator
representation has been proposed in \cite{JKKS2} for the three-point
function of non-BPS operators.}

The paper is organised as follows.  In the rest of this section we
present the setup and the results.  Section \ref{section:2} gives
formulation of the octagon as a Fredholm pfaffian.  Section
\ref{section:perturbativeoctagon} is devoted to the perturbative
analysis.  Section \ref{section:CFTrepresentation} gives the CFT
fermionic representation of the octagon.

 \subsection{The simplest 4-point correlation function}
 
A half-BPS field is characterised by its position $ x$ in the
Minkowski space, a null vector $y$ giving its $\sosix$ polarisation,
and its scaling dimension $K$,
\be 
\mathcal{O}_i = \Tr [(y_i\cdot \Phi(x_i))^{K}].
\ee

\begin{figure}
         \centering
       	 \begin{minipage}[t]{0.8\linewidth}
            \centering
            \includegraphics[width=7.2 cm]{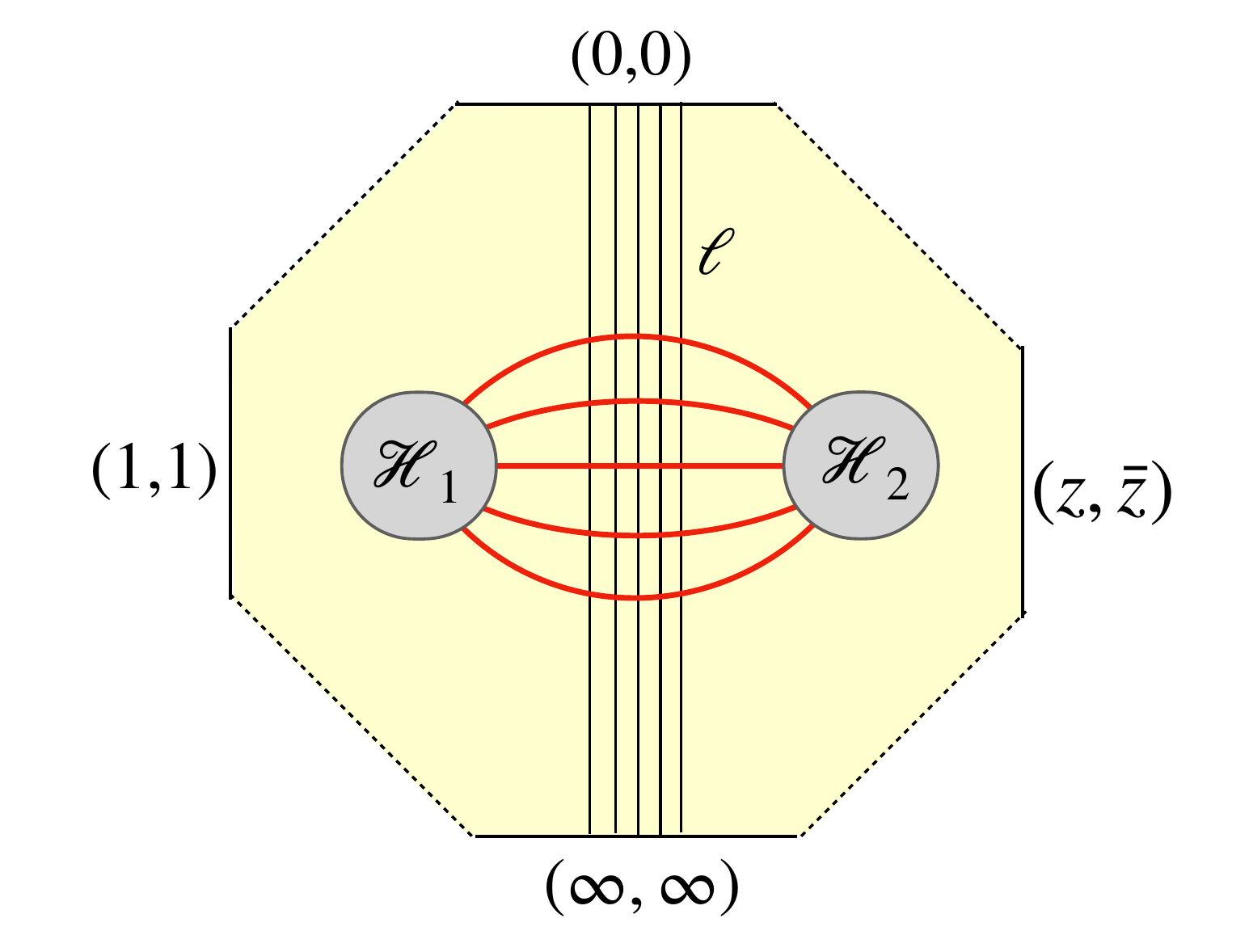}
   \caption{ \small A sketch of the octagon $\mathbb{O}_\ell$.  The
   parameters $z$ and $\bar z$ fix the conformal geometry of the
   octagon.  The red lines symbolise the mirror particles propagating
   between the two hexagons, each one characterised by a rapidity $u$
   and a bound state number $a$.  Each mirror particle has to pass
   across a `bridge' composed of $\ell$ Wick contractions.  }
  \label{fig:Octagon} 
         \end{minipage} 
            \end{figure}

The correlation function for four such operators can be computed
non-perturbatively by the hexagonalisation method designed first in
\cite{BKV1} for the computation of the three-point functions and
adjusted for the four-point functions in
\cite{Fleury:2016ykk,Eden:2016xvg, Fleury:2017eph}.  The sum over
virtual states prescribed by the hexagonalisation simplifies for heavy
fields (large $K$) and particular choices of the polarisations because
some of the channels of propagation of the virtual particles are
suppressed.  In the correlators analysed in \cite{Coronado:2018ypq}
the sum over virtual particles factorises into a sum of products of
two octagons.  An octagon $\mathbb{O}_\ell(g, x, \bar z, \alpha, \bar\alpha)$,
sketched in Fig.  \ref{fig:Octagon}, is composed of two hexagons glued
together by inserting a complete set of virtual states.  It depends
only on the bridge length $\ell$ between the two hexagons, 't Hooft
coupling $g$ and the cross ratios \re{crossratios}
 in the coordinate and in the flavour spaces
  \be
  \begin{aligned}
  \la{crossratios} z\bar z &= {x_{12}^2 x_{34}^2\over x_{13}^2
  x_{24}^2}=u,\ \qquad \qquad\qquad (1-z)(1-\bar z)= {x_{14}^2
  x_{23}^2\over x_{13}^2 x_{24}^2}=v, \\
  \alpha\bar\alpha &= {(y_1 \cdot y_2) (y_3 \cdot y_4)\over (y_1 \cdot y_3)
  (y_2 \cdot y_4)}, \ \qquad (1-\alpha)(1-\bar\alpha) = {(y_1 \cdot y_4) (y_2
  \cdot y_3)\over (y_1 \cdot y_3) (y_2 \cdot y_4)} .
    \end{aligned}
  \ee

 The simplest four-point function that leads to such a factorisation,
 named in \cite{Coronado:2018ypq} the {\it simplest}, is characterised
 by $(y_1\cdot y_4)=(y_2\cdot y_3)=0$.  For example (the dots stand
 for the sum over permutations)
\be
\la{simplest4pf}
\begin{aligned}
\mathcal{O}_1 (0) &= \tr ( Z^{K\over 2} \bar X^{K\over 2}) +...  , \qquad
\quad \mathcal{O}_2 (z, \bar z)= \tr (X^K) , \\
\mathcal{O}_4 (\infty)&= \tr ( Z^{K\over 2} \bar X^{K\over 2}) +...,
\qquad\quad \quad \quad \mathcal{O}_3 (1) = \tr(\bar Z^K) \, .
\end{aligned}
\ee
In the limit $K\to \infty$ the {\it simplest} four-point correlator
factorises to a product of two identical octagons with $\ell=0$, and
$\a=\bar \alpha=1$,
\be \langle \mathcal{O}_1\mathcal{O}_2\mathcal{O}_3\mathcal{O}_4\rangle \ \ \underset{K\to\infty} =\
{\left[ \mathbb{O}_{0} (z,\bar z, 1, 1)\right] ^2 \over (x_{12}^2 x_{34}^2 x_{13}^2
x_{24}^2 )^{K/2}}.  \ee
There is another class of four-point functions considered in
\cite{Coronado:2018ypq}, which are expressed in terms of octagons
$\mathbb{O}_\ell $ with $\ell>0$.  In the recent paper
\cite{Bargheer:2019kxb} the non-planar four-point correlators are
expressed as polynomials of $\mathbb{O}^2$.  It this paper we will focus
exclusively on the octagon form factor $\mathbb{O}_\ell(z,\bar z, \alpha, \bar\alpha)$.

\subsection{The octagon form factor}

 Here we remind the series expansion of the octagon as a sum over
 mirror particles mainly with the purpose of introducing our notations
 which are slightly different than those of \cite{Coronado:2018ypq}.
 As usual we denote by $u$ the rapidity of the particles in the 2d
 quantum field theory describing the dynamics of the single-trace
 operators in $\CN=4$ SYM. The dispersion relation for the physical
 particles $ E_a^2= {a^2\over 4} + 4 g^2 \sin^2{p_a\over 2} $ is
 parametrised by the \Zh map
\begin{align}\la{paramux}
{u\over g} = x+{1\over x}, \qquad x(u)={u\over g} \( {1\over 2} +
{1\over 2}\sqrt{1- {4 g^2\over u^2}}\).
\end{align}
 The momentum and the energy of the physical particles of type $a$ are
 given by
\begin{align}
p_a(u) = - i (\ID^a - \ID^{-a})\log x, \quad E_a(u) =- {ig\over 2} \(
\ID^a - \ID^{-a}\) (x- {1\over x})\, ,
\end{align}
where $\ID= e^{i \partial_u/2}$ is
 the shift operator,  to be used repeatedly in the following,
 \be
 \la{defID}
 \mathbb{D}:   f(u) \to  f(u+ i/2).
 \ee
 For some computations this operator representation of functions with
 shifted arguments can be quite efficient.  We will use sometimes the
 commonly accepted notations,
 \be
 \la{shifts}
 f(u\pm i a/2)=f^{[\pm a]}(u)=  \ID^{\pm a}f (u) = f(u) ^{\ID^{\pm a}}.
 \ee

We are interested in summing over the particles in the mirror dynamics
whose energy and momentum are given by
\begin{align}
 \tilde p_a(u) 
 =   \textstyle{1\over 2}   g
 \( \ID^a + \ID^{-a} \) (x-{1\over x}),
\quad  \tilde E_a(u) =   (\ID^a+\ID^{-a})\log x  .
\end{align}

 We consider an octagon with four physical and four mirror edges with
 the corresponding BMN vacuum at each physical edge, as shown
 schematically in Fig.  \ref{fig:Octagon}.  The octagon is obtained by
 gluing the hexagons $\mathcal{H}_1$ and $\mathcal{H}_2$ along the common edge
 $(0,0)$--$(\infty, \infty)$ by inserting a complete set of virtual
 states $\psi$ with energies $ {\tilde E}_\psi$.  A state $\psi$ contains an
 arbitrary number of fundamental particles and their bound states
 transforming in the skew-symmetric representations of
 $\psu(2|2){\times }\psu(2|2) $.  Symbolically
 \be \mathbb{O}_\ell = \sum_\psi \langle  \mathcal{H}_2|\psi\rangle  \,
 \, e^{-
 {\tilde E}_\psi \ell} \, \langle  \psi|\mathcal{H}_1\rangle .  
 \ee
 To write the explicit expression one should bring the two hexagon
 operators to the canonical hexagon $\mathcal{H}$.  The dependence on the
 cross ratios in the coordinate and flavour spaces appears through the
 similarity transformations $\mathcal{H}_1\to \mathcal{H}$ and $\mathcal{H}_2\to\mathcal{H}$:
  \be \la{octagonconceptually} \mathbb{O}_\ell
   (z, \bar z, \alpha, \bar \alpha) 
  =
  \sum_\psi \langle \mathcal{H}|\psi\rangle  
  e^{- \tilde E_\psi \ell} \, e^{2i \tilde p_\psi \xi} \, e^{i L_\psi
  \phi}\, e^{i R_\psi \theta} \, e^{i J_\psi \varphi}\langle  \psi|\mathcal{H}\rangle .  \ee
The parameters $\phi ,\xi, \theta, \varphi $ conjugated to $\mathbf{L
}, \bm{p}, \mathbf{R }, \mathbf{J }$ are related to the cross ratios
in the Minkowski and in the flavour spaces, eq.  \re{crossratios}, as
\be \la{angles} \begin{aligned} z &= {e^{-\xi+i\phi } } , \quad\quad
\bar z = {e^{-\xi - i\phi} } , \\
 \alpha& = { e^{\varphi -\xi  + i\theta}  } ,
  \ \ \
  \bar \alpha = { e^{\varphi -\xi - i\theta}} .
 \end{aligned}\ee

An $n$-particle virtual state $\psi$ is completely characterised by
the rapidities and the bound state numbers $(u_j, a_j)$ of the
individual particles $(j=1,..., n)$.  Taking into account the explicit
form of the hexagon form factors, one writes \re{octagonconceptually}
as the following series of multiple integrals,
 \be
 \begin{aligned}
  \la{defIn} \mathbb{O}_{\ell}&= \sum_{n=0}^\infty {1\over n!} \sum_{a_1,...,
  a_n\ge 1} \int \prod_{j=1}^n {du_ j \over 2\pi }\ (-1)^{a_j} \
  \bm{\mu} _{a_j}(u_j, \ell, z, \bar z)\ W^{\text{matrix}}_{a_1...  a_n}
  \prod_{j<k}^n \tilde H_{a_j, a_k} (u_j, u_k) \ .
 \end{aligned}
 \ee
 The three factors in the integrand/summand are defined as follows.

$\bullet$ The local integration measure
  \be\la{defmubar}   \bm{\mu} _{a}(u)=
   \tilde \mu_{a}(u) \ e^{-\tilde E_{a}(u) \ell } \ e^{2i \xi \,
   \tilde p_{a}(u) } \ee
contains the intrinsic measure coming from the canonical hexagons
  \be\la{measuremua} \tilde \mu_a(u)= { x^{[+a]} - x^{[-a]} \over {
  x^{[+a]} x^{[-a]} -1} } \times {1\over i g} \ \prod_{\varepsilon=\pm} { 1
  \over\(x^{[\varepsilon a]} -1/x^{[\varepsilon a]} \) } \ee
 as well as factors depending on $\xi = -  \textstyle{1\over 2}  \log z\bar z$ and the
 bridge length $\ell$.
   
 $\bullet$ The symmetric bilocal factors $\tilde H_{ab}(u,v)$ are
 produced by the diagonal part of the hexagon
 weights in mirror-mirror kinematics  
\be
\begin{aligned} \la{bilocal}
\tilde H_{ab}(u,v)
= \prod_{\varepsilon, \d =\pm}{x^{[\varepsilon a]} -y^{[\d b]}\over x^{[\varepsilon a]} y^{[\d
b]}-1}\
,
\end{aligned}\ee
 where $x=x(u)$ and $y=x(v)$.  The bi-local weights can be written
 with the help of the shift operator \re{defID} as
\be
\begin{aligned} \la{bilocalK}
\tilde H_{ab}(u,v)&=K(u,v)^{(\ID_u^a+\ID_u^{-a})(\ID_v^b+\ID_v^{-b})}
, \qquad K(u,v)= {x-y\over xy-1} .
  \end{aligned}\ee

$\bullet$ The  factor $ W^{\text{matrix}}_{a_1...  a_n}
$ accounts for the matrix part.  The
contraction of the flavour indices in the matrix factor can be done
using the unitarity of the $\psu(2|2)$ S-matrix and the result is
expressed in terms of the traces (transfer matrices), which in absence
of physical excitations are simply the $\psu(2|2)$
characters.\footnote{This is true only if the integrand does not
contain singularities as e.g. double poles.} The characters are
expressed in terms of the variables $ \phi, \varphi, \alpha$ parametrizing
the cross sections in the coordinate and in the flavour space.  As
suggested in \cite{Fleury:2016ykk,Fleury:2017eph} and elaborated upon
in \cite{Coronado:2018ypq}, one should consider two ways of dressing
the mirror basis with $\CZ$ markers and then take the average between
the two choices.  The matrix part takes the form of a sum of two terms
 \be
 W^{\text{matrix}}_{a_1... a_n} 
 =  {1\over 2}  \(  \prod_{j=1}^n  \chi_{a_j}  ^+ 
 + \prod_{j=1}^n  \chi_{a_j}  ^-\)
 \ee
  where $\chi_a^\pm $ are the characters of the antisymmetric
  representations of $\psu(2|2)$, \cite{Fleury:2016ykk,Fleury:2017eph}
  (cf eq.  (47) of \cite{Fleury:2016ykk})
 \be\begin{aligned} \la{su22charakiri} \tr_a[(-1)^F \ e^{ \varphi
 \mathbf{J }+i \phi \tilde \LL + i \theta \tilde \RR} ] &= \chi_a ^+ +
 \chi_a ^-,\qquad \chi_a ^\pm &= (-1)^a \ { \l^{\pm} \ {\sin (a\phi
 )\over \sin\phi }} ,
\end{aligned} 
\ee
 where the dependence on $\varphi$ and $\theta$ is carried out by the
 factor\footnote{The factor $\l^\pm$ is related to $\CX^\pm$ in
 \cite{Coronado:2018ypq} by $ \mathcal{X}^+=-\frac{(z-\alpha )
 \left(\bar{z}-\alpha \right)}{\alpha } = \l^+ e^{-\xi};\
 \mathcal{X}^-=-\frac{\left(z-\bar{\alpha }\right)
 \left(\bar{z}-\bar{\alpha }\right)}{\bar{\alpha }}= \l^- e^{-\xi}$.
 }
  \be \la{deflpm} \l^\pm(\phi, \varphi,\theta) = 2\left[  \cos \phi -\cosh (
  \varphi \pm i \theta )\right] 
 .
\ee
This form of the $\psu(2|2)$ characters follows from the generating
function
 \be\begin{aligned}\la{diffoTa} \mathcal{W}_\pm (t) & = \sum_{a=0}^\infty
 (-1)^a \chi_a ^\pm \, e^{a t} = {(1- e^{ \pm \varphi + i \theta} e^{t}
 )(1-e^{\mp \varphi- i\theta} \, e^t) \over(1- e^{i\phi }\, e^t)(1- e^{-
 i\phi } \, e^t)} \\
 &
= 1+{\l^\pm \over 2 ( \cosh t -\cos \phi ) }.  \end{aligned} \ee

The final expression of the series expansion for the octagon takes a
form resembling a Coulomb gas of dipole charges
 \be
 \begin{aligned}
 \la{defOctserer} \mathbb{O}_{\ell}(g , z, \bar z, \alpha, \bar\alpha)
 &=1+\sum_{n=1}^{\infty} \ \CX_n(\phi, \varphi, \alpha) 
 \ \CI_{n, \ell}(z, \bar z)
\end{aligned}
 \ee
 with 
 \be 
 \CI_{n, \ell}(z, \bar z)= 
  {1\over n!} \,e^{n \xi}
  \sum_{a_1,..., a_n\ge 1}
  \prod_{j=1}^n 
  {\sin (a_j\phi )\over \sin\phi } 
   \int
 \prod_{j=1}^n {du_ j \over 2\pi }\
{ \bm{\mu} }_{a_j}(u_j) 
     \prod_{j<k} \tilde
    H_{a_j, a_k} (u_j, u_k)\,,  
    \ee
\be \CX_n = {(\l^+)^n+(\l^-)^n\over 2}\ e^{-n \xi}\,.  \ee
  The multiple contour integrals were evaluated by residues order by
  order up to $n=4$ in \cite{Coronado:2018ypq}.  The perturbative result for the
  octagon obtained in \cite{Coronado:2018ypq} matched the five loops
  results in \cite{Chicherin:2018avq} obtained previously using the
  conformal symmetry, the hidden dual conformal symmetry and analytic
  bootstrap conditions.  As explained before, it allows to extend
  these results to any loop order.

 \subsection{Summary of the  results  }

 \subsubsection{The octagon$^2$  as a determinant}

 Based on the representation of the series \re{defOctserer} as
 Fredholm pfaffian, explained in section \ref{subsec:fredhpfaff}, we
 give an explicit formula for the octagon in terms of the square root
 of the determinant,
   \be\begin{aligned} \la{PfOctbispfa0} \mathbb{O}_{\ell}(z, \bar z, \alpha,
   \bar\alpha)& & =  \textstyle{1\over 2}  \sum_{\pm} \sqrt{ \Det\left[  \mathbf{I} - \l^\pm \ \C
   \KKK\right] } .
              \end{aligned}
    \ee
 The matrices $ \mathbf{I}, \C$ and $\KKK$ are semi-infinite matrices
 of the type $\mathbf{M}= \{\mathrm{M}_{m,n}\}_{m,n=0}^\infty$.  The
 matrices $ \mathbf{I} $ and $\C$ are standard,
 \be \la{defC} \mathrm{I}_{m,n} = \d_{m,n}\ , \quad
 \mathrm{C}_{nm}=\d_{n+1, m }- \d_{n, m+1}, \qquad m,n\ge 0, \ee
while the matrix $ \KKK $ depends on the gauge coupling $g$ and the
  cross ratios in the coordinate space through the angle $\phi$ and
the parameter $\xi$ defined in \re{angles}.  Its matrix elements
$\K_{nm}$ are given for any coupling $g$ by a non-singular integral of
a product of two Bessel functions,
  \be\la{integralforK0}
\begin{aligned}
\K _{mn} &= { g \over 2 i } \int _{ |\xi|}^\infty d t { \( i\sqrt{
t+\xi \over t-\xi}\)^{ m-n }\!\!\!\!  - \( i\sqrt{ t+\xi \over
t-\xi}\)^{ n-m } \over \cos\phi - \cosh t }
 \ J_{m+\ell}(2g\sqrt{ t^2-\xi^2})\ J_{n+\ell}(2g\sqrt{ t^2-\xi^2})
  ,
  \end{aligned} 
 \ee 
while the dependence on the cross ratios for the polarisations occurs
through the factors $\l^\pm$ defined in \re{deflpm}.  According to
\re{PfOctbispfa0} the octagon is expressed in terms of the moments of
the matrix $\C\KKK$,
  \be
  \la{Octexpa}
 \begin{aligned}
 \mathbb{O}_{\ell} &=
    \textstyle{1\over 2}  \, e^{ \CS ^+} +  \textstyle{1\over 2}  \, e^{ \CS ^-}, \\
 \CS ^\pm& = -  \textstyle{1\over 2}  \sum_{n=1}^\infty {( \l^\pm)^{n}\over n} \tr [(\C
 \KKK)^n].
    \end{aligned}
    \ee
 The $n$-th moment gives the contribution of a connected cluster of
 $n$ virtual particles.  Formula \re{Octexpa} provides an explicit
 expression for the Coronado integrals in \re{defOctserer} in terms of
 the moments of the matrix $\C\KKK$
 \be
 n \CI_{n,\ell}= -{1\over 2}\sum_{k=0}^{n-1} \CI_{k, \ell}  \, \tr [(e^\xi \C \KKK)^{n-k}]\,, n\ge 1\,, \   \CI_{0,\ell}=1\,.
 \ee
In the main body of the paper we also  represent \re{PfOctbispfa0}
as the pfaffian of a semi-infinite matrix.

   \subsubsection{Perturbative expansion}
   
 The expression \re{PfOctbispfa0} is true for any coupling $g$.  If
 one is interested in the perturbative expansion at weak coupling, the
 formula can be made more explicit.  Within the perturbative expansion
 the octagon in \re{PfOctbispfa0} becomes a determinant and the
 formula for the octagon reads
\be \la{octagonfinalpR0} \mathbb{O}_\ell = \textstyle{1\over 2}  \sum_{\pm} \det(1+ \CX^\pm
\CRR) ,
\ee
where $\CX^\pm = \l^\pm e^{-\xi}$ and the elements of the matrix
$\CRR=\{ \CR_{k,j}\}_{k,j\ge 0}$ are infinite linear combinations of
the ladder integrals \re{deffk},
\be
\begin{aligned}
\la{seriesR} &\CR_{k j} =\!\!\!\!\!  \sum_{ p= \max ( k +j+\ell, 1+j+\ell)}^\infty
(-1)^{p-\ell} \
\frac{ \Gamma(2 p) \big( 2p(2 k +\ell) \ - \delta _{k ,0} \ (p - j )
(p+j+\ell)\big)}{ \prod\limits_{  \pm} \Gamma( p   \pm (k -
j)+1) \ \Gamma(p   \pm (k +j+\ell)+1) }\ f_p(z, \bar z) \ g^{2p} \ ,
 \end{aligned}
\ee
where $f_p$ are given in \re{deffk}.  In actual computations it is
convenient to truncate the semi-infinite matrix $\CRR$ to a $N\times
N$ matrix $\CRR_{N\times N}= \{\CR_{k,j}\}_{0\le k,j \le N-1}$.  Such
a truncation reproduces the perturbative expansion of the octagon to
loop order $2N-1$,

\be\la{octagonfinalpR0bis} \mathbb{O}_\ell = \textstyle{1\over 2}  \sum_{\pm} \det(1+ \CX^\pm\
\CRR)_{_{N\times N}} + o(g^{^{4N+2\ell}}).  \ee
 For example, the truncation to a $3\times 3$ matrix gives the
 perturbative expansion up to $o(g^{12})$,
\be\begin{aligned}\no \mathbb{O}_{\ell=0}&= \textstyle{1\over 2}  \det\( 1+\CX ^+\CRR\)_{3\times
3}+  \textstyle{1\over 2}  \det\( 1+\CX ^-\CRR\)_{3\times 3} + o(g^{12})\\
   =1 &+ \CX_1 \ \( f_1 g^2-f_2 g^4 +  \textstyle{1\over 2}  f_3 g^6 -\textstyle{5\over
   36} f_4 g^8 + \textstyle{7\over 288} f_5 g^{10}
\) 
\\
 &+ \CX_2 \ \( \textstyle{1\over 12} (f_1 f_3-f_2^2)g^8-
 \textstyle{1\over 24} (f_1 f_4 -f_2 f_3) g^{10}
\) \ + o(g^{12}).  \end{aligned}\ee
The nine loop result presented in \cite{Coronado:2018ypq} is
reproduced by truncating to a $5\times 5$ matrix.  To compare with
\cite{Coronado:2018ypq} one should take $f_n= n!(n-1)!  F_n$.

 \subsubsection{The null-square  limit}
 
 Here we comment shortly the null-square limit when the intervals
 between two subsequent operators in Fig.  1 become light-like, $
 x_{12}^2, x_{13}^2, x_{34}^2, x_{42}^2 \to 0$, or $ z\to 0, \ 1/\bar
 z\to 0$.  In \cite{Coronado:2018cxj} it was conjectured that the
 logarithm of the octagon enjoys the following simple asymptotics,
\be
\begin{aligned}
\la{lcasym}
\lim\limits_{^{z\to0}_{ \bar z\to\infty}} \log\mathbb{O}&=
-  \tilde\Gamma(g) \(\log(-z) + \log(-1/\bar z)\)^2 
\\
&+ {1\over 2} g^2 \( [\log(-z)]^2 + [\log(-1/\bar
z)]^2\)+\text{constant}.  
\end{aligned}
\ee
The coefficient $\tilde\G$ is somehow similar to the cusp anomalous dimension.

We were able to reproduce this asymptotics in the weak coupling limit
from the determinant formula.  In the null-square limit $\CX_n = \bar
z^n$ and the determinant formula \re{octagonfinalpR0} gives for the
logarithm of the octagon
   \be \la{logoctt} \log\mathbb{O} = \Tr \log (1+\bar z \CRR)= \bar z \, \tr
   \CRR - {\bar z^2\over 2} \tr \CRR^2 + {\bar z^3 \over 3} \tr \CRR^3
   -...  \ee
  The matrix elements of  $\CRR$ simplify in this limit and
  for the determinant we obtain, after massive cancellations, the
  asymptotics \re{lcasym}. We checked that the perturbative expansion
  of $\tilde\Gamma(g)$ is in agreement with \cite{Coronado:2018cxj}.
  To compute $\tilde\Gamma(g)$ to $N$ loops it is sufficient to keep
  the first $N$ terms in the series \re{logoctt} and truncate 
    $\CRR$ to an $N\times N$ matrix.  For example, if we take
  $N=10$, we obtain the nine-loop result
\be\la{tildeGamma}
\begin{aligned}
  \tilde \Gamma(g)&= \frac{1}{2} g^2-\frac{ 1}{6} \pi^2
  g^4+\frac{8}{45} \pi^4 g^6 -\frac{68 }{315} \pi^6g^8+ \frac{3968
  }{14175} \pi^8g^{10} - \frac{176896 }{467775} \pi^{10}g^{12}\\
 &+\frac{22368256 }{42567525}
 \pi^{12}  g^{14}
 -\frac{475939328 }{638512875}   \pi^{14 }  g^{16}
 +\frac{104932671488 }{97692469875}    \pi^{16 }  g^{18}  
 + o(g^{20}) .
\end{aligned}\ee

\section{The octagon as a   pfaffian }
\label{section:2}

\subsection{The octagon as a Fredholm pfaffian }
\label{subsec:fredhpfaff}
 
It has been pointed out in \cite{Basso:2017khq} that in some simple
configurations the sum over the virtual particles can be recast in the
form of a Fredholm pfaffian \cite{2000math......6097R}.  In this
subsection we will develop the pfaffian representation in full detail.
We first remind the definition of a Fredholm pfaffian in general.
 
 \medskip

 Let $\KK(u_1,u_2) $ be a $2\times 2$
antisymmetric matrix kernel of trace class defined on $\IR$,
 \be \la{2x2matK} \KK(u_1,u_2)= \left(\begin{array}{cc}K^{++}(u_1,u_2)
 & K^{+-}(u_1,u_2) \\ K^{-+}(u_1,u_2) & K^{--}(u_1,u_2)
 \end{array}\right), \quad K^{\varepsilon_1 \varepsilon_2}(u_1,u_2) = - K
 ^{\varepsilon_2\varepsilon_1}(u_2, u_1).  \ee
 Denote by $\JJ$ the standard $2\times 2$ antisymmetric matrix 
 kernel      
 \be
 \la{defIJ}
\begin{aligned}
 \qquad \JJ= \left(\begin{array}{cc}0 & 1 \\ -1 & 0
 \end{array}\right)\ \d(u_1-u_2) .
\end{aligned}
\ee
 The Fredholm pfaffian $\pf [\JJ+\KK]$ is defined as the series of
 multiple integrals with some integration measure $d\mu(u)$
 \be \la{Fpfaffexp} \pf [\JJ+\KK]= \sum_{n=0}^\infty {1\over
 n!}\int\limits _\IR\prod_{j=1}^n d\mu(u_j) \, \pf
 [\KK_n(u_1,...,u_n)] , \ee
where $\KK_n(u_1,...,u_n)$ is the $2n\times 2n$ 
antisymmetric matrix
\be \la{deffinmKK} \KK_n(u_1,...,u_n)= [\KK(u_j,u_k)]_{1\le j,k\le n}
\ee
where $ \KK(u,v)$ is the $2 \times 2 $ matrix \re{2x2matK} .  The
relation with the Fredholm determinant is
 \be\la{PfDet}
 \pf [\JJ+\KK]=\sqrt{\det\left[  \II - \JJ  \KK\right] }
 \ee
where $\II$ is the $2\times 2$ matrix kernel of the identity operator.

\medskip

Now let us focus on the series \re{defOctserer}.  Using the pfaffian
version of the Cauchy identity
\be \la{PfCauchi} \prod_{j<k}^{2n} {x_j-x_k\over x_jx_k-1}=\pf\(
\left[ {x_j- x_k\over { x_jx_k} -1}\right] _{i,j=1}^{2n}\) \ee
we can write the product of the bi-local weights \re{bilocalK} as a
pfaffian.  In order to illustrate how the series is assembled into a
Fredholm pfaffian, let us first neglect the bound states with $a \ge
2$ and consider only the sum over fundamental particles ($a=1$).  It
is quite obvious that the grand canonical sum of virtual particles is
the expansion of a Fredholm pfaffian of the skew-symmetric kernel
 \be\begin{aligned} \la{Kpmdef} K^{\varepsilon_1,\varepsilon_2}(u_1,u_2) &= K(u_1 + i
 {\varepsilon_1\over 2} , u_2 + i {\varepsilon_2\over 2}) ,\qquad \varepsilon_1,\varepsilon_2 = \pm.
 \end{aligned}
 \ee
which is obtained by shifting with $\pm i/2$ the arguments of the
kernel \re{bilocalK},
\be \la{defK} K(u,v)={x(u) - x(v) \over x(u) x(v) - 1}.  \ee
In order to formulate the exact statement we need to specify the
measure.

The $n$-particle pfaffians \re{Fpfaffexp} then give the product of the
bi-local factors in the integrand in \re{defOctserer} times pieces of
the local measure.  For that we will compare the terms with $n=1$ and
$n=2$ in the expansion \re{Fpfaffexp} with the contributions of one-
and two-particle states to the octagon.

$\bullet$ For $n=1$
  \be
 \KK_1(u)=
 \left(
\begin{array}{cc}
 0 & \frac{\xp-\xm}{\xm\xp-1} \\
 \frac{\xm-\xp}{\xm \xp-1} & 0 \\
\end{array}
 \right) \ee 
 and the pfaffian gives the first factor in the intrinsic measure
 \re{measuremua} with $a=1$,
\be \quad \pf( u)\equiv \pf\left[  \KK_1(u)\right]  = {\xp-\xm \over \xp\xm-1}.
\ee

 $\bullet$ For $n=2$    
  \be
 \KK_2(u_1,u_2)=
 \left(
\begin{array}{cccc}
 0 & \frac{{\xp}-{\xm}}{{\xm} {\xp}-1} & \frac{{\xp}-{\yp}}{{\xp}
 {\yp}-1} & \frac{{\xp}-{\ym}}{{\xp} {\ym}-1} \\
 \frac{{\xm}-{\xp}}{{\xm} {\xp}-1} & 0 & \frac{{\xm}-{\yp}}{{\xm}
 {\yp}-1} & \frac{{\xm}-{\ym}}{{\xm} {\ym}-1} \\
 \frac{{\yp}-{\xp}}{{\xp} {\yp}-1} & \frac{{\yp}-{\xm}}{{\xm} {\yp}-1}
 & 0 & \frac{{\yp}-{\ym}}{{\ym} {\yp}-1} \\
 \frac{{\ym}-{\xp}}{{\xp} {\ym}-1} & \frac{{\ym}-{\xm}}{{\xm} {\ym}-1}
 & \frac{{\ym}-{\yp}}{{\ym} {\yp}-1} & 0 \\
\end{array}
 \right) ,\  \ x=x(u_1)\,, y=y(u_2) \,.
 \ee
It is identified with the matrix in \re{PfCauchi} for $n=2$, with
$x_1=x^+\,, x_2=x^-\,, x_3=y^+\,, x_4=y^-$.

The pfaffian gives the bilocal weight $\tilde H_{11}(u_1,u_2)$ in the
integrand for the octagon, eq.  \re{bilocal}, times the factors
$\pf(u_1)\pf(u_2)$,
 \be\begin{aligned}
 &\pf(u_1,u_2)\equiv \pf  \left[  \KK_2(u_1,u_2)\right]  
  &= \pf(u_1)\pf(u_2) \tilde H_{11}(u_1,u_2).
 \end{aligned} 
\ee

 $\bullet$ For general $n$ the pfaffian of $\KK_n$ is given by the
 product
 \be \pf(u_1, ..., u_n)\equiv \pf \left[  \KK_n(u_1,..., u_n)\right] =
 \prod_{j=1}^n \pf(u_j) \prod_{j<k}^n \tilde H_{11}(u_j , u_k).  \ee
 Hence the measure $d\mu(u)$ is given by the measure \re{measuremua}
 stripped from the first factor.

Now let us consider the complete sum.  In order to include the bound
states, we have upgraded the continuous variable $u\in\IR $ to a pair
$(u,a)\in \IR\times \IN$.  Instead of the $2\times 2$ matrix kernel
$\KK (u_1, u_2)$ defined on $\IR^{\times 2}$, eq.  \re{Kpmdef}, we
have to deal with the $2\times 2$ matrix kernel $\KK (u_1,a_1;
a_2,u_2)$ defined on $ (\IR\times \IN)^{\times 2}$ and having matrix
elements
  \be \la{Kab} K^{\varepsilon_1, \varepsilon_2}(u_1, a_1; u_2,a_2) = K(u_1 ^{[\varepsilon_1
  a_1]} , u_2^{[\varepsilon_2 a_2]}),\quad \varepsilon_{1,2}=\pm\ .  \ee
 The functions $x^{[\pm a]} =x(u\pm i a/2)$ are taken in the first
 sheet, where $|x(u)|>1$.  The pfaffians for one- and two-particle
 states are
 \be
 \la{pfaffs}
 \begin{aligned}
 \pf(u,a) &\equiv \pf [\KK_1(u,a)] = {x^{[+a]}-x^{[-a]} \over
 x^{[+a]}x^{[-a]}-1} \\
\pf(u,a; v,b) &\equiv  \pf[\KK_2(u,a; v,b)]=  \pf(u,a)\pf(v,b)\
H_{ab}(u,v) .
\end{aligned} 
\ee
The pfaffian for a single particle gives a piece of the measure
\re{measuremua}.  Let us  denote the remaining piece by $\o(u,a)$:
 \be
 \la{defoa}
 \tilde \mu_a(u) =
 \pf(u,a) \,  \o(u,a) ,
  \quad 
 \o(u,a) = {1\over i g} \
 \prod_{\varepsilon=\pm}    { 1
 \over\(x^{[\varepsilon a]}  -1/x^{[\varepsilon a]} \)  }.
\ee
 Now we are able to formulate the exact claim.  The expansion of the
 octagon \re{defOctserer} as a sum of two Fredholm pfaffians
 \be
 \begin{aligned}
 \la{defInPf} \mathbb{O}_{\ell} &={1\over 2} \sum_{\pm} \sum_{n=0}^\infty
 {(\l^\pm)^n\over n!} \sum_{a_1,...,a_n\ge 1} \int \prod_{j=1}^n \
 d\mu(u_j,a_j) \ \pf\left[  \KK_n(u_1, a_1;...  ; u_n, a_n)\right] .
 \end{aligned} \ee
with the   integration measure given by
\be\la{defmuua} d\mu(u, a) = {\sin(a\phi)\over \sin\phi}\, {du\over
2\pi} \ \o(u,a)\, \ e^{-\tilde E_{a}(u) \ell } \ e^{2i\xi \, \tilde
p_{a}(u) } .  \ee
In view of the subsequent analysis, 
it will be important that the integration measure
\re{defmuua}  can be 
written  in a factorised form 
\be
\la{measuredma}
\begin{aligned}
d\mu(u,a)&= {1\over ig   } \, {\sin(a\phi)\over \sin\phi}\, 
\  {du\over 2\pi} \ \O_\ell (u+ia/2) \  \O_\ell   (u-ia/2) 
   ,
  \end{aligned}
   \ee
  with the function $\O_\ell(u)$ defined as
  \be
  \la{defOl}
   \begin{aligned}
    \O_\ell (u ) &\equiv {e^{ig \xi\, [x(u)-1/x(u)]}\over x(u)-1/x(u)}
    \ x(u)^{-\ell} .  \end{aligned} \ee

To summarise, the expansion \re{defOctserer} of the octagon sums up to
a Fredholm pfaffian which by \re{PfDet} is a square root of a Fredholm
determinant,
 \be\la{OctPfDet}
 \begin{aligned}
  \mathbb{O}_{\ell}(z, \bar z, \alpha, \bar\alpha)&=  \textstyle{1\over 2}  \sum_{\pm} \  \pf\( \JJ+\l^\pm
  \KK\) \\
 &=  \textstyle{1\over 2}   \sum_{\pm} \  \sqrt{\Det(\II - \l^{\pm}  \JJ  \KK)}
 ,
 \end{aligned}
 \ee
with the matrix elements of the kernel $\KK$ defined in \re{Kab} and the integration measure
given by \re{measuredma}-\re{defOl}.

\subsection{ Summing up the bound states}
\label{sec:sumbs}
 
In order to perform the sum over the bound states it is more
convenient to start with the determinant representation in the second
line of \re{OctPfDet},
  \be
  \la{Octexp}
 \begin{aligned}
 \mathbb{O}_{\ell}(z, \bar z, \alpha, \bar\alpha)&=
   \textstyle{1\over 2}   \,  e^{  \CS ^+} +    \textstyle{1\over 2}   \,  e^{  \CS ^-},
  \end{aligned}
 \ee
 with
 \be
  \la{defCA} \begin{aligned} \CS ^\pm &= \textstyle{1\over 2}  \Tr\log ( \II-\l^\pm
 \JJ \KK) \\
  &= -  \textstyle{1\over 2}  \sum_{n=1}^\infty {\( \l^\pm\)^{n}\over n}{1\over (ig )^n}
  \sum_{a_1, ..., a_n\ge 1} \sum_{\varepsilon_1, ...  , \varepsilon_n= \pm} \
  \prod_{j=1}^n {\sin(a_j\phi)\over \sin\phi} \\
 &\times
  \       \int _{\IR}  \prod_{j=1}^n {du_j\over 2\pi }\ 
    \O_\ell^{ [- \varepsilon_j a_j] }(u_j)\  \O_\ell^{ [+ \varepsilon_j a_j] }(u_j)
       \\
  & \times \varepsilon_1 \, K(u_1 ^{[ - \varepsilon_1 a_1]} , u_2 ^{[ \varepsilon_2 a_2]} ) \ \
  \varepsilon_2 \, K(u_2 ^{[ - \varepsilon_2 a_2]} , u_3 ^{[ \varepsilon_3 a_3]} ) \ ...  \
  \varepsilon_n \ K(u_n^{[ - \varepsilon_n a_n]} , u_1^{[ \varepsilon_1 a_1]} ) .
    \end{aligned}
    \ee
     %
  The factors $\varepsilon_j$ come from $J ^{\varepsilon, -\varepsilon} = \varepsilon, \ J^{\varepsilon,\varepsilon} =
  0, \ \varepsilon = \pm $. 
       
    We will take advantage of the fact that the elements of the matrix
    kernel \re{Kab} are obtained by shifting the arguments of the
    scalar kernel \re{defK}.  This will allow us to replace the sum
    over the labels $a_j$ by an appropriate difference operator.  This
    is not unrelated to the fact that the sum over the $\psu(2|2)$
    characters gives the generating function \re{diffoTa}.  We will
    obtain the same generating function, but with $t$ replaced by $i
    \partial_u$.

   First we notice that the integrand is analytic in the strip $-
   a_j/2< \text{Im}(u_j) < a_j/2$, which allows us to displace the
   contour for $u_j$ by $ -i \varepsilon_j (a_j-\e) /2 $.  Here $\e$ is a
   small positive quantity which will be sent to 0 in the final
   expression.  We will keep it finite during the computation because
   it indicates whether we are above or below the real axis on which
   $x(u)$ has a cut.  The displacement of the contours can be
   compensated by shifts of the variables $u_j\to u_j + i \varepsilon_ja_j
   /2$.  As a result of this manipulation the integrand for the $n$-th
   term takes the form
\be \la{defCAshifted} \begin{aligned} & \int _{\IR} {du_j\over 2\pi }\
\O^{[- i \epsilon \varepsilon_j]}(u_j )\O^{[ 2 \varepsilon_j a_j]}(u_j) \\
   &\times \varepsilon_1 \ K(u_1^{[-i\epsilon \varepsilon_1]}, u_2^{[2 \varepsilon_2 a_2]}) \ \
   \varepsilon_2 \ K( u_2^{[-i\epsilon \varepsilon_2]}, u_3^{[ 2 \varepsilon_3 a_3]} ) \ ...  \
   \varepsilon_n \ K( u_n^{[-i\epsilon \varepsilon_n]}, u_1^{[ 2 \varepsilon_1 a_1]} ) \\
   &= \int _{\IR+ i \epsilon \varepsilon_j} {du_j\over 2\pi }\ \ \varepsilon_1 \ \hat
   K(u_1^{[-i\epsilon \varepsilon_1]}, u_2^{[2 \varepsilon_2 a_2]}) \ \ \varepsilon_2 \ \hat K(
   u_2^{[-i\epsilon \varepsilon_2]}, u_3^{[ 2 \varepsilon_3 a_3]} ) \ ...  \ \varepsilon_n \ \hat K(
   u_n^{[-i\epsilon \varepsilon_n]}, u_1^{[ 2 \varepsilon_1 a_1]} ),
    \end{aligned}
    \ee
  where the non-flat part of the measure is absorbed in the kernel by
  dressing it with two factors \re{defOl},
  \be
  \la{defhatK}
   \begin{aligned}
  K(u, v)&\ \to\ \hat K(u, v)= \O_\ell(u ) \, K( u, v ) \ \O_\ell(v) .
   \end{aligned} 
    \ee
 The advantage of this rewriting is that now the multiple sums over the
 bound state labels $a_1,..., a_n$ in the cyclic integral decouples.
 The sum over each label $a_j$ can be most easily performed by
 introducing the shift operator $\ID_{u_j}$ with the result being a
 difference operator acting on $\hat K(u_{j-1}, u_j)$.

In this way we traded the multiple sum over the bound state labels for
a more complicated $2\times 2$ matrix kernel
$ \IK =\{ \IK^{\e,\d}\}_{\e,\d=\pm}$ whose matrix elements are  
 \be \begin{aligned} \la{KBb} \IK ^{\varepsilon,\d}(u,v)& \defeq {1\over ig }
 \sum_{a\ge 1} {e^{i a \phi } - e^{- i a \phi } \over e^{i \phi } -
 e^{- i \phi } } \hat K(u + i \varepsilon \epsilon , v+ i \d a)
    \\
&=-{1\over   ig  }
\
  { 1
 \over 
  2\cos\phi  -
 2\cos ( \partial_v)  
 }    \hat  K(u+ i\varepsilon \epsilon , v+i \d \epsilon ) 
 .
\end{aligned} \ee
   The action of the operator function in the last line is well
   defined for any $\e>0$.  Returning to the series \re{defCA}, we
   write it as the logarithm of a Fredholm determinant of the $2\times
   2$ matrix kernel $\JJ\IK$
  \be \la{defCAb} \begin{aligned} \CS ^\pm &= - \sum_{n=1}^\infty {
  (\l^\pm)^{n}\over 2n} \sum_{\varepsilon_1, ...  , \varepsilon_n= \pm} \int _{\IR}
  \prod_{j=1}^n {du_j\over 2\pi } \prod_{j=1}^n \varepsilon_j \ \IK^{-\varepsilon_j,
  \varepsilon_{j+1}}(u_j, u_{j+1}) \\
 &=  \textstyle{1\over 2}  \Tr \log(\II -\l^\pm   \JJ \IK),
    \end{aligned}
    \ee
   and respectively for the octagon  
   \be \la{DetOct} \mathbb{O}_{\ell}(z, \bar z, \alpha, \bar\alpha)=  \textstyle{1\over 2} \sum_{\pm}
   \sqrt{ \Det\left[  \II-\l^\pm \JJ \IK \right] } .  \ee
The corresponding pfaffian representation is
  \be \la{PfffOct} \mathbb{O}_{\ell}(z, \bar z, \alpha, \bar\alpha)=  \textstyle{1\over 2} \sum_{\pm}
  \pf\left[  \JJ+\l^\pm \IK
 \right] 
   .
   \ee

 \subsection{From Fredholm kernel to a semi-infinite matrix}
 
 \la{sec:Frtosemiinf} Our next goal is to perform the multiple cyclic
 integrations.  This can be done by breaking the cyclic integral into
 a sum of products of independent integrals.  The formula we are going
 to obtain is an infinite-dimensional version of the pfaffian
 integration theorem \cite{Akemann:2007wa,1751-8121-40-36-F01}.  The
 idea is simple.  We will expand the scalar kernel \re{defK} in a
 complete set of harmonic functions on the \Zh Riemann surface, which
 will allow us to disentangle the multiple integrals in \re{defCAb}
 and represent the Fredholm pfaffian as the Pfafian of a semi-infinite
 skew-symmetric matrix.
   
The mode expansion  of the scalar  kernel $K$ reads
for $ |x|>1$ and $|y|>1 $
\be\la{basisK} K(u,v)= \frac{x-y}{x y-1}= \sum _{m, n=0}^{\infty }
x^{-n} \ \mathrm{C}_{nm} \ y^{-m} , \ee
where  
 \be \la{defCbis} \mathrm{C}_{nm}=\d_{n+1, m }- \d_{n, m+1}, \qquad
 m,n\ge 0.  \ee
We will denote by $\C$ the semi-infinite antisymmetric matrix   
with elements $\mathrm{C}_{nm}$.
Assuming that the variables $u$ and $v$ are in the first sheet
($|x|>1, |y|>1$), we represent the dressed kernel $\hat K$, eq.
\re{defhatK}, as a double series
\be
  \hat K(u,v) 
=
\sum_{n,m\ge 0} \mathrm{C}_{nm} 
\
{ \O_{\ell + n}(u)
\
\O_{\ell + m}(v)}\, ,\ \ \ \ 
\ee
with the functions $\O_j (u)$ defined by \re{defOl}.  The kernel
\re{KBb} expands in a similar way,
 \be
\begin{aligned}
\IK^{\varepsilon \d}(u,v) & =-{ 1\over ig} \sum_{m,n\ge 0} \O_{\ell +
n}(u+\varepsilon i \epsilon ) \ { \mathrm{C}_{nm} \over 2 \cos\phi - 2\cos \partial_v }
\O_{\ell+m}(v +\d i \e)
.
    \end{aligned}\ee
The operator sandwiched between the two $\O$-functions is obtained as
in \re{KBb}.  The only subtlety is that when $\d=-$, it represents a
power series of $\ID^{-1}$ which shifts by $- i/2$.

After expanding in this way the kernels in the $n$-th term of the
series \re{defCAb}, the chain of $n$ entangled integrals decouple into
a sum of products of $n$ simple integrals.  The exponents $\CS ^\pm$
take the form
  \be \la{defCAbb} \begin{aligned} &\CS ^\pm= -  \textstyle{1\over 2}  \sum_{n=1}^\infty
  {( \l^\pm)^{n}\over n} \ \tr (\C \KKK)^n
    \end{aligned}
    \ee
 where $ \KKK$ is a semi-infinite skew-symmetric matrix with matrix
 elements ($m,n\ge 0$)
 \be \la{defKker}
\begin{aligned}
  \K  _{mn}&=  \CP_{m n}-\CP_{n m}\,\\
\CP_{mn} &=-  {1\over 2 ig } 
\int {du\over 2\pi }  \ \O _{\ell+n}(u -  i 0)
   { 1 
 \over \cos\phi - \cos \partial_u } \ \O _{\ell+m}(u+ i 0) .
 \end{aligned}\ee

 The above procedure is nothing but a change of the basis for the
 operator representing the Fredholm kernel.  The determinantal
 representation \re{DetOct} takes in the new discrete basis the form
   \be\begin{aligned} \la{DetOctbis} \mathbb{O}_{\ell}(z, \bar z, \alpha,
   \bar\alpha)&=  \textstyle{1\over 2} \sum_{\pm} \sqrt{ \Det\left[  1 - \l^\pm \C \KKK\right] }
    \end{aligned}
    \ee
    and the corresponding pfaffian formula is\footnote{If we
    approximate the Fredholm kernel with a degenerate one obtained by
    truncating the sum in \re{basisK} to $0\le m,n\le N-1$, then
    $\pf[\C^{-1}]= (-1)^N$.  }
   \be\begin{aligned} \la{PfOctbispfa} \mathbb{O}_{\ell}(z, \bar z, \alpha,
   \bar\alpha)&=  \textstyle{1\over 2} \sum_{\pm} {
 \pf\left[    \C^{-1} - \l^\pm    \KKK\right] 
 \over \pf\left[    \C^{-1}  \right] }  
              \end{aligned}.
    \ee
with
\be
\left[   \C^{-1}\right] _{kl}=
\sum_{n\ge 0}\(\delta _{2 n+1+(-1)^k k+l}-\delta _{2 n+1+k+(-1)^l l}
\).  \ee

\subsection{Integral representation for the matrix elements}
\la{sec:integralrep}

The operator expression \re{defKker} for the elements of the matrix
$\KKK$ can be transformed into an integral by a Fourier
transformation,
\be\la{Kmndef}
\begin{aligned}
 \CP_{mn}&=   {1\over  i g} \sum_{a=1}^\infty 
{ \sin a\phi \over \sin \phi}\int_{-\infty}^{\infty} \, dt \,
\tilde{\Omega}_{\ell+n}^{-}(t) \, e^{-at}\, \tilde{\Omega}^+_{\ell+m}(-t)
\\
 &=  { 
1 \over ig} \int _{-\infty}^\infty d t \; {\tilde \O^{- }_{\ell+n}(
t)\; \tilde\O^{ +}_{\ell+m}( -t) \over \cosh t- \cos\phi } ,
 \end{aligned}
\ee
where $\tilde \O^\pm _j(t)$ is the Fourier image of $\O_j(u\pm i \e)$,
 \be \begin{aligned} \tilde\O^{\pm }_k( \mp t)&= \int_{-\infty}^\infty
 {du\over 2\pi}\, e^{\mp i t u }\ \O_k(u\pm i 0) \\
    &=  g    \int_{\IR\pm i 0}
 {dx(u ) \over 2\pi}
     \ {e^{ \mp i  g(x \t_{\pm}   + {1\over x}\t_ {\mp})}  \over    x^{k+1}   }  
       ,
   \ \ \  \t_\pm 
   = t\mp \xi =t\pm \log |z|.
   \end{aligned}
   \ee

These integrals are computed by residues,
   \be\begin{aligned} \la{fOm} \tilde\O_k^{\pm}(\mp t) &=\mp \alpha(
   \t_{\pm })ig \({i} \sqrt{\t_- \over \t_+}\)^{ \mp k } \
    J_k(2  g \sqrt{\t_+\t_-}). \qquad   
     \end{aligned}
      \ee
The Heaviside functions $\alpha(\t_\pm)$ give a restriction of the lower
bound in the integral in \re{Kmndef} to $t= |\xi|$.  Hence the
representation \re{integralforK0} of the matrix as an integral of a
product of Bessel functions, which we write again for convenience,
  \be\la{integralforK}
\begin{aligned}
 \K _{mn} &= { g \over 2 i } \int _{ |\xi|}^\infty d t { \( i\sqrt{
 t+\xi \over t-\xi}\)^{ m-n }\!\!\!\!  - \( i\sqrt{ t+\xi \over
 t-\xi}\)^{ n-m } \over \cos\phi - \cosh t } \ J_{m+\ell}(2g\sqrt{
 t^2-\xi^2})\ J_{n+\ell}(2g\sqrt{ t^2-\xi^2}) ,
  \end{aligned} 
 \ee 
This completes the derivation of our main formula \re{PfOctbispfa0}.

  \section{The  perturbative octagon }
  \label{section:perturbativeoctagon}
 
Throughout this section we will assume that we are in the weak
coupling regime.  Within the perturbative expansion we are going to
render the representation \re{PfOctbispfa0}-\re{integralforK0} more
explicit and show that it matches the weak coupling results in
\cite{Coronado:2018ypq, Coronado:2018cxj}.  For that we need to
compute the perturbative expansion for the matrix elements $\K_{mn}$.
   
  \subsection{ Weak coupling expansion of the  matrix $\KKK$
 }

In \cite{Coronado:2018ypq} it is claimed that the dependence of the
octagon on the positions and on the polarisations of the four
operators is only through the $j$-loop ladder integrals, which are
evaluated as linear combinations of polylogarithms
\cite{USSYUKINA1993136}\footnote{ In the notations of
\cite{USSYUKINA1993136}, $F_j(z,\bar z)= -v \Phi^{(j)}(u,v)$.}
 \be \begin{aligned} \la{defF} F_{j}(z, \bar z) & = \sum_{s=0}^j {
 2j-s\choose j } \
 { | \log z\bar z |^{s} \over s!} 
\ {\mathrm{Li}_{2j-s}(z) - \mathrm{Li}_{2j-s}(\bar z)\over z-\bar z} =
 {1\over j!(j-1)!} f_j(z, \bar z) .
 \\
 \end{aligned}
 \ee
 In Appendix \ref{AppendixA} we obtain for the matrix elements
 \re{integralforK0} a double expansion in ladder integrals $f_k$ {\it
 and} the parameter $\xi = - \log\sqrt{z\bar z}$,
  \be\la{A911}
\begin{aligned}
  \K_{ m+r,m}   
& = e^{-\xi} \sum_{p=0}^\infty { 2\ell+2m+2p+r\choose p} {(-1)^{l+m}
(ig)^{ 2\ell +2m+2p+r+1} \over ( \ell+m+p)!( \ell+m+p+r)!} \\
& \times \sum_{k=1}^{[{r+1\over 2}]} \, {r-k\choose k-1} |2
\xi|^{r-2k+1} 
\, f_{ \ell+m+p+k}(z,\bar z) \qquad (r\ge 1).
\end{aligned}
\ee
 The terms with nontrivial powers of $ \xi$ are not consistent with
 the crossing invariance of the 4-point function.  Fortunately it
 turns out that they do not show up in the traces $\tr [ (\C\KKK)^n]$.
 The redundant $\xi$-dependence of the expansion \re{A911} seems to be an
 artefact of the choice of the basis \re{basisK}.  That is, there is a
 rotation $O \C\KKK O^{-1} $ which kills all the dependence on $\xi$
 of the matrix elements, while preserving all traces.  Assuming that
 this is the case, we replace the series for $\K_{n,m}$ with the
 truncated series obtained by removing all terms with non-trivial
 powers of $\xi$.
  \footnote{Note that setting $\xi=0$ in the ladder functions as well,  simplifies the integral formula  \re{integralforK0} and in particular imposes
  the opposite parity of the indices $m=n+1$ mod 2. }
   
 Let us denote the truncated series by $\K^\circ_{n,m}$.  For even
 values of $r$ all powers of $\xi$ in \re{A911} are nontrivial and we
 get \be \la{KcircE} 
 \K^\circ_{m+2s,m}=0 .  \ee
 For odd powers of $r $ we denote $r=2s-1$ with $s\ge 1$.  Retaining
 from the sum in the second line of \re{A911} only the term with $k=
 (r+1)/2 = s$ we get
  \be\la{Kcirc}
\begin{aligned}
 \K^\circ_{m+2s-1,m}
= e^{-\xi}\ \sum_{j=l+m+s}^{\infty} {2j-1\choose j-l-m-s}
{(-1)^{j-l-m} \over (j-s)!(j+s-1)!} g^{2j} f_j(z,\bar z).
\end{aligned}
\ee
Since $\tr [(\C\KKK)^n] = \tr[(\C \KKK^\circ)^n]$, in order to obtain the 
perturbative series of the octagon up to
$g^{4N}$ it is sufficient to replace the semi-infinite matrix matrix
$\KKK^\circ$ to a ${2N\times 2N}$ matrix
\be \la{defKKKcutoff} \KKK^\circ_{_{2N\times 2N}} = \( \K^\circ
_{ij}\)_{i,j=0, 1, ..., 2N-1}.  \ee
The truncated determinant formula \re{DetOctbis} reproduces the
perturbative series for the octagon up to order $g^{4N}$,
   \be\begin{aligned} \la{PfOctbispfaNN} \mathbb{O}_{\ell} &=  \textstyle{1\over 2} \sum_{\pm}
   \sqrt{ \Det\left[  1 - \l^\pm \C_{_{2N\times 2N}} \KKK_{_{2N\times
   2N}}\right] } + o(g^{4N+2\ell }) \\
&   =  (-1)^N  \textstyle{1\over 2} \sum_{\pm} 
  \pf\left[    \C^{-1}_{_{2N\times 2N}}-  \l^\pm   \  \KKK^\circ_{_{2N\times 2N}}\right]   
    + o(g^{4N+2\ell})\  .
              \end{aligned}
    \ee

\subsection{The perturbative octagon as a determinant}

The formula \re{PfOctbispfaNN} can be made aesthetically and
conceptually more attractive if we repackage the $2N\times 2N$
matrices $\C_{_{2N\times 2N}}$ and $\KKK_{_{2N\times 2N}}$ in
\re{PfOctbispfaNN} as the $N\times N$ matrices
\be \begin{aligned} 
\la{defCC}
 \CCC_{_{N\times N}}&= \big( \CC_{jk} \big)_{j,k=0, N-1},\qquad \CC_{jk}
 \equiv \mathrm{C}_{2j, 2k+1} = \d_{j-k }- \d_{j-k-1} \, 
 \\
 \CKK_{_{N\times N}} &= \big(  \CK_{ij }\big)_{i,j=0,..., N-1},
\qquad \CK_{ij }  =   \K^\circ_{2i+1, 2j} .
 \end{aligned}\ee
Then  due to   \re{KcircE}  we have 
$\tr [(\C\KKK^\circ)^n] =2 \, {\rm tr}[(\CCC \CKK)^n]$ and for any ``cutoff'' $N$ 
the following  identity holds,
\be  \la{2NvsN}
 \begin{aligned}
{ \ \ (-1)^N } \pf\left[  ( \C^{-1}- \l\ \KKK^\circ) _{_{2N\times 2N}}\right] 
 =   \det\left[   (1-\l\   \, \CCC \CKK )_{_{N\times N}}\right]    .
 \end{aligned} \ee
 The determinant formula for the perturbative octagon now takes the
 form\footnote{For  $\xi =0 $  the determinant formula extends to  any  $g$ with $\KKK^\circ$  replaced by the (simplified) integral representation 
\re{integralforK0}\,.}
\be\la{octagonfinalpR} \mathbb{O}_\ell = \textstyle{1\over 2}  \sum_{\pm} \det(1+ \l^\pm
e^{-\xi}\ \CRR)_{_{N\times N}} + o(g^{^{4N+2\ell}}),
\ee
where the matrix elements of $\CRR =-e^\xi\ \CCC\CKK$ are given by a
linear combination of ladders with rational coefficients,
  \be \la{seriesRbis} {\begin{aligned} \CR_{ij}(z, \bar z, g)&= \sum
  _{p=\max(i+j+\ell, 1+j +\ell)}^{2N} \CR_{ij}(p)\ f_p(z,\bar z) \ g^{2p}, \\
   \CR_{ij}(p)
& = \frac{(-1)^{p-\ell} (2 p-1)!  (2 p (2 i+\ell) (1-\d_{i,0}) -\delta
_{i,0} (p-j-\ell) (j+p))}{(i-j+p)!(-i+j+p)!  (-i-j-\ell+p)!
(i+j+\ell+p)!}.
 \end{aligned}  
 }
 \ee

\subsection{Expansion to  nine loops }

In order to compare with the nine-loop result obtained in
\cite{Coronado:2018ypq} by direct computation of the multiple
integrals in the octagon series, we need to expand the determinant in
\re{octagonfinalpR} for $\ell=0$  with $N=5$,
{\small
 \be\begin{aligned} &\mathbb{O}_{\ell=0}= \textstyle{1\over 2}  \det\( 1+\l ^+
e^{-\xi} \CRR\)_{5\times 5}+  \textstyle{1\over 2}  \det\( 1+\l ^-\ e^{-\xi}
\CRR\)_{5\times 5} \\
 &=\ 1+ \CX_1 \( f_1 g^2-f_2 g^4 +  \textstyle{1\over 2}  f_3 g^6 -\textstyle{5\over 36}
 f_4 g^8 + \textstyle{7\over 288} f_5 g^{10} -\frac{7}{2400} f_6
 g^{12} +\textstyle{\frac{11 }{43200} f_7 }g^{14} \ + \ ...  \)
\\
 &\qquad \ + \CX_2 \ \( \textstyle {f_1 f_3-f_2^2\over 12} g^8- {f_1
 f_4 -f_2 f_3\over 24} g^{10}+ {7 f_1 f_5-9 f_3^2+2 f_2 f_4\over 720}
 g^{12} + \frac{3 f_3 f_4-2 f_2 f_5-f_1 f_6}{720} g^{14}+...  \) \\
 &\qquad \ + \CX_3 \( \textstyle{f_1 f_5 f_3-f_3^3+2 f_2 f_4 f_3-f_1
 f_4^2-f_2^2 f_5\over 34560} g^{18}
\)\ + o(g^{20} ),
\end{aligned}
\ee
}
with
\be
 \CX_n = {\l_+^n+\l_-^n\over 2}\ e^{-n \xi},
 \ee
which matches  completely the result obtained in   \cite{Coronado:2018ypq}.

  \subsection{Fishnets}

 In \cite{Coronado:2018cxj}, the octagon was expanded in a basis of
 minors of the matrix \re{detf}.  In particular, the lowest loop order
 $n$-particle contribution is proportional to the determinant of the
 matrix \re{detf} restricted to the first $n$ rows and columns, which
 has been identified in \cite{Basso:2017jwq} with the Feynman integral
 for an $n\times n$ fishnet,
 \be
 \begin{aligned}
\la{sumfishnets} \mathbb{O}_{\ell=0} = \sum_{N=0}^\infty \CX_N \ g^{2N^2}\
\(C_N + o(g^2)\) \, , \qquad C_N = {{\det\(\left[ f_{i+j+1} \]_{i,j=0,...,
N-1}\)\over \prod_{i=0}^{N-1} (2i)!(2i+1)!}} \, .
\end{aligned}
 \ee
  This property of the octagon is obvious from our determinant
  representation \re{octagonfinalpR}, which can be written as a sum
  over minors of the matrix $\CRR$, eq.  \re{seriesR},
 \be\begin{aligned} \la{minoresR} \mathbb{O}_{\ell=0}&= \sum_{N=0}^\infty \CX_N
 \sum_{^{0\le i_1< ...<i_N} _{0\le j_1 <...<j_N}} \det \(\left[  \CR_{i_\alpha
 j_\b}\]_{\alpha,\b=1,..., N}\) \\
&= \sum_{N=0}^\infty 
\CX_N\
\(
  \det\CR_{_{N\times N} } \, + o(g^2)\) .
\end{aligned}
 \ee
 Indeed, to the lowest order the determinant of the matrix
 $\CR_{_{N\times N} } $ is given by the fishnet integral normalised as
 in \re{sumfishnets},
  \be
 \det \CR_{_{N\times N} }  =C_N\ 
  g^{2N^2}+o(g^{2N^2+2}).
  \ee

\subsection{The null-square  limit}

Here we check that the logarithm of the octagon in the light cone
limit $z\to 0, 1/ \bar z\to 0$ takes the form \re{lcasym} claimed in
\cite{Coronado:2018cxj}.  We reproduce this behavior from our
determinant formula \re{octagonfinalpR}.  The light cone limit
corresponds to large imaginary values of the angle $\phi$.  It is
convenient to parametrise $z$ and $\bar z$ with the variable $y= i(\pi
- \phi)\to\infty$,
 \be
 z  =-e^{- \xi -y},\bar z = - e^{- \xi+y}.
 \ee
 The light cone limit corresponds to $y\to +\infty$.  Furthermore in
 the light cone limit $\l^\pm =- e^{y}$, $\CX_n = (-1)^n
 e^{ny}e^{-n\xi} = \bar z^n$, and the logarithm of the octagon is \be
 \la{explogO} \log\mathbb{O} _{\ell=0}= \tr \log (1+\bar z \CRR)= \bar z \,
 \tr \CRR - { \bar z ^2\over 2} \tr \CRR^2 + { \bar z ^3\over 3} \tr
 \CRR^3 -...  \ .  \ee

 Take the matrix $\CR$ for $\ell=0$, 
\be
\begin{aligned}
\la{seriesR0} &\CR_{ij} = \sum_{ p= \max ( i+j , 1)}^\infty (-1)^{p }
(2 p-1)!
\frac{ 4 i p \ - \ (p - j ) (p+j )\ \delta _{i,0} }{ \prod\limits_{\varepsilon
= \pm} ( p + \varepsilon (i- j))!  ( p + \varepsilon (i+j ))!  }\ f_p \ g^{2p},
\end{aligned}\ee
   and substitute the expansion of the ladders in the light cone limit
   \cite{USSYUKINA1993136}
\be\begin{aligned} \la{expansionf} -v f_p(z,\bar z)& = \sum _{m=0}^p \sum _{n=0}^p {p \choose m}{p\choose
n} \left(1-2^{1- m-n }\right) {1+
(-1)^{m+n}\over p} \\
&\times  ( m+n)!  \
\zeta (m+n)\ ( y+\xi )^{p-m}( y-\xi)^{p-n}  \
\end{aligned}\ee
where $-v=-(1-z)(1-\bar z) \to  \bar z= -e^{y+\xi} $.
E.g.
\be f_1=\xi ^2-y^2-\frac{\pi ^2}{3}, \ \ f_2 = -\frac{1}{2}
\left(y^2-\xi ^2\right)^2+\frac{1}{3} \pi ^2 \left(\xi ^2-3
y^2\right)-\frac{7 \pi ^4}{30} \ee
 etc. In general $\bar z f_n$ is a polynomial in $y$ and $\xi$ of degree $2n$.
 
If we want to evaluate $\tilde \G(g) $ to $N$ loops, we compute the
expansion of the first $2N$ terms of the series \re{explogO} with the
matrix $\CR$ truncated to $N\times N$,
   \be \log\mathbb{O} = \bar z \, \tr \CR_{_{N\times N}} - {\bar z^2\over 2}
   \tr \CR_{_{N\times N}} ^2 + {\bar z^3 \over 3} \tr \CR_{_{N\times
   N}} ^3 -...  +{\bar z^{2N}\over 2N}\CR_{_{N\times N}} ^{2N} +
   o(g^{2N}).  \ee
 Taking $N=10$ we find, with the help of Mathematica, an expression
 which is of the  form      
 \re{lcasym}
\be\begin{aligned}
\log\mathbb{O}&=
 -4 y^2\ \tilde\Gamma(g) + g^2 \(y^2 + \xi^2\) + \text{const},
 \end{aligned}\ee
with   $\tilde \G$ given by  
\re{tildeGamma}. 

%
%

\section{CFT representation  }
\label{section:CFTrepresentation}

 An important property of the sum over virtual particles used to
 solder the two hexagons into an octagon, eq.  \re{defIn}, is that the
 scattering of each particle with the other particles is taken into
 account by a local weight factor.  As a consequence, the two-particle
 interaction is symmetric and the grand canonical sum can be mapped
 onto a Coulomb gas of dipole charges.  We will show that this Coulomb
 gas appears as the expansion of a certain expectation value in a
 theory of a twisted chiral boson defined on the \Zh plane.  The
 twisted boson changes sign every time it circles one of the two
 branch points at $u= \pm 2 g$.  We will also show that the vertex
 operators of the bosonic field can be mapped to bilinears of real
 fermions.

 \subsection{The octagon in terms of a free   chiral boson}

 We introduce a chiral gaussian field $\varphi(u)$ defined on the \Zh
 plane.  The field is defined in terms of a mode expansion over a
 complete set of functions $ \psi_n(u) = x(u)^n$ $(n\ge 1)$, where $x=
 x(u) $ is the \Zh map $u \to x(u) $,
  \be \la{gfield} \varphi(u) = \hat q+ \hat p\ln x(u) -\sum_{n\ne 0}
  {J_{n}\over n} x(u)^{-n} , \ee
\be \la{ccr} [ J_n, J_m] =n\delta_{m+n, 0}; \ \ [\hat p,\hat q] = 1.
\ee
 The Fock space is build on the 
vacuum states $\langle 0 | $ and $ |0\rangle  $ with the properties
 \be
\begin{aligned}
&J_n |0\rangle  =0 , \ \ (n>0); \ \ \ \hat p |0\rangle  = 0 , \\
&\langle 0 |  J_n  =0 , \ \  (n<0); \ \ \  \langle 0 | \hat q =
 0\, ,
 \end{aligned}
  \ee 
  so that the two-point function is
\be
\la{propphi}
 \langle \varphi(u)\varphi(v)\rangle  = \ln(x(u)-x(v))
 , \qquad  
  |x|>|y| >1
 .
\ee

Let us denote by $\varphi^{(\varepsilon)}(u)\, , \varepsilon=\pm 1\,,$ the value of the
gaussian field on each of the two sheets of the Riemann surface with a
cut $[-2g, 2g]$.  The mode decomposition for $\varphi^{(-)}(u)$ is given
by \re{gfield} with $x\to 1/x$.  The 2-point functions of the pair of
fields $\varphi^{(\varepsilon)}(u)$ are given by
 \be \la{propphii} \langle
\varphi^{(\varepsilon_1)}(u)\varphi^{(\varepsilon_2)}(v)\rangle = \ln(x^{\varepsilon_1}(u)-x^{\varepsilon_2}(v))
.  \ee 
The monodromy around a branch point exchanges $\varphi^{(+)}(u)$
and $\varphi^{(-)}(u)$ and is diagonalised by 
\be \phi(u)
={\varphi^{(+)}(u)-\varphi^{(-)}(u)\over \sqrt{2}}, \quad \tilde \phi(u) =
{\varphi^{(+)}(u)+\varphi^{(-)}(u)\over \sqrt{2}}.  \ee 
We will call the field
$\phi$, antisymmetric with respect to the inversion $x\to 1/x$,
twisted and the field $\tilde \phi$ untwisted.  The two-point function
of the twisted field is
 \def\rg{{\color{red}\g}} \be \la{twistedfield} 
 \langle 0|
 \phi(u)\phi(v)|0\rangle =  \textstyle{1\over 2}  \ln { (x-y) ({1\over x}-{1\over
 y})\over (x-{1\over y})({1\over x}-y)} = \ln{ x-y\over xy-1} \ee
 while the two-point function of the untwisted field is
  \be 
  \langle 0|
 \tilde \phi(u)\tilde \phi(v)|0\rangle =  \textstyle{1\over 2}  \ln {
 (x-y)(\textstyle{1\over x}-\textstyle{1\over y}) (x-\textstyle{1\over
 y})(\textstyle{1\over x}-y)} = \ln(\textstyle{{u-v\over g}}) .  \ee
 The two-point function of the normal ordered exponents of the twisted
 field reproduces the antisymmetric kernel \re{defK}
\be
:e^{\phi(u)} :\ :e^{\phi(v)}: =   
K(u,v) :e^{\phi(u) +\phi(v)}:
\ee 
  The factor \re{bilocal} is reproduced by \be\la{haboper}
  H_{ab}(u,v) = \langle 0| :e^{\phi(u+ ia/2)+ \phi(u-ia/2)}: \
  :e^{\phi(v+ ib/2)+ \phi(v-ib/2)}: |0\rangle.  \ee

For any operator made out of the oscillators \re{ccr} we define
another vacuum expectation value as
\be \la{newexp} \langle \mathcal{O}\rangle = \langle 0| \mathcal{O} e^{-i \sqrt{2} g
\xi J_{-1} -{\ell\over \sqrt{2}} \, q} |0\rangle \ee
 so that 
 \be \langle 
\phi(u) \rangle = \phi_c(u)=ig\xi (x- {1/x})-\ell \ln x\,.  \ \ee
 Furthermore define the operator\footnote{The shift of the gaussian
 field seems somewhat {\it ad hoc}.  It can be obtained by defining
 the vertex operator without normal ordering and computing the
 singular part by point splitting by $\varepsilon$,
$$
 \la{expvalue} \langle 0 | e^{\phi (u)}|0\rangle _{\e}= e^{ {1\over 2} \langle 0 | \phi
 (u+ \e) \phi (u)|0\rangle }=\({\e\ x'(u)\over x^2-1}\)^{1/2} ={
 \sqrt{\e}/\sqrt{g}\over x(1-1/x^2)} = \sqrt{ \e\over g } {g\over
 \sqrt{u^2- 4 g^2}}.
 $$
 }
\be \Phib (u)=\log {g \over \sqrt{u^2- 4 g^2}}\ +\phi(u) \ee so that
expectation value \re {newexp} of the vertex operator is exactly the
function $\O_\ell(u)$ defined in \re{defOl}, \be\la{expvalue1} \langle 
:e^{\Phib (u)} :\rangle  = \O_\ell(u) = e^{\Phib_c(u)} \ \ee while the
two-point function reproduces the scalar kernel \re{defK},
 \begin{align}
\la{twopfK} \langle  :e^{\Phib(u)}: : e^{\Phib(v)}: \rangle &=
\O_\ell(u)\O_\ell(v)\ { x(u)-x(v) \over x(u) x(v)-1 } = \hat K(u,v).
\end{align}
The integration measure \re{defmubar} is given by
\be\begin{aligned} \bm{\mu} _a(u ) &= {1\over i g} \hat K(u+ ia/2,u-
ia/2)\,.  \end{aligned}\ee
The series \re{defOctserer} is obtained by expanding the following
expectation value,
 \be
 \la{OctaOper}
 \begin{aligned}
 & \mathbb{O}_{\ell}=  \textstyle{1\over 2} \sum_{\pm} \left\langle \exp\left[  \l^\pm\ \int {du
 \over 2\pi i g } \ : e^{ \Phib (u-i0)}: { 1 \over 2\cos\phi - 2 \cos
 \partial_u } : e^{ \Phib (u+i0)} :
 \]
\right\rangle\, .
  \end{aligned}
 \ee

\subsection{Real fermion  }

The exponential field 
\be \la{fermions} \Psib (u) = :e^{\phi (u)} : \ee
behaves as a real fermion.  The correlation function of $2n $ such
fermions is the pfaffian of the two-point functions
 \begin{align}
 \la{tpffer} \langle \Psib(u_1) ...  \Psib(u_{2n})\rangle  &=
\pf\(
\left[ {x(u_j)- x(u_k)\over { x(u_j)x(u_k)} -1}\]_{i,j=1}^{2n}\)
\end{align}
   The series \re{defCA} gives the free energy of these fermions.
   Thefree energy is the sum of all fermionic loops (the factor
   $1/2$ reflects the fact that the fermionic lines are non-oriented).

The analytic field $\Psib(u)$  is completely determined by the mode expansion  
\be\la{discretecor}
\Psib(u) = \sum _{m\ge 0} \Psib_m \ x(u)^{-m},
\quad 
\langle 0| \Psib_m\Psib_n|0\rangle  = \mathrm{C}_{mn}.
\ee
where $C_{mn} $ is given by \re{defC} .  Indeed, we have
\be \langle0| \Psib(u)\Psib(v)|0\rangle = {x- y\over xy-1} =
\sum_{m,n\ge 0} x^{-m}C_{mn}\, y^{-n} .  \ee
The quadratic form in the exponent becomes in the discrete basis
\be {1\over 2 i g} \int {du \over 2\pi } \ \Psib(u-i 0) e^{ \Phi_c
(u-i0)} { 1 \over \cos \phi - \cos \partial_u } e^{ \Phi_c (u+i 0)}
\Psib(u+i 0) =-{1\over 2} \sum_{m,n=0}^\infty \Psib_m \K_{nm }\Psib_n
\ee
 with $\K_{mn}$ given by \re{defKker}.
  In terms of the discrete fermionic modes the operator expression of
  the octagon reproduces the pfaffian representation of the octagon
  \re{PfOctbispfa},
 \be
 \la{fermion-discrete}
 \begin{aligned}
 \mathbb{O}_{\ell}=  \textstyle{1\over 2} \sum_{\pm}  \langle 0 | e^{ - {{\l^\pm}\over
 2} \sum_{m,n\ge 0} \Psib_m \K _{nm} \Psib_n }  |0  \rangle
 =  \textstyle{1\over 2} \sum_{\pm} { \pf\left[  \C^{-1} - \l^\pm \KKK\]
   \over \pf\left[  \C^{-1} \] } \, .
  \end{aligned}
 \ee
If the expectation value is understood as an integral over real
grassmanian variables, then the first term of the pfaffian comes from
the matrix of the quadratic form which produces the correlator
\re{discretecor}.

 \section{Conclusions}

  In this paper we give an explicit expression for the octagon form
  factor, which is the building block in the construction of the class
  of four-point functions of heavy half-BPS operators considered in
  \cite{Coronado:2018ypq}.  As it was recently discovered in
  \cite{Bargheer:2019kxb}, the higher terms in the $1/N_c$ expansion
  of these functions can be expressed in terms of the octagon as well.

 Our main result is the determinant formula for the octagon form
 factor, eq.  \re{DetOctbis}, for any value of the gauge coupling.  In
 this way the computation of the octagon is reduced to that of a
 single integral, eq.  \re{integralforK0}.  The structure of this
 integral reminds the one for the generating function for the ladder
 Feynman integrals \cite{Broadhurst:2010ds}.\footnote{We thank L.
 Dixon for bringing to our attention ref.  \cite{Broadhurst:2010ds}.}
 At weak coupling the traces \re{defCAbb} computed with the integral
 in question are linear combinations of products of ladder integrals.
 \footnote{The effective truncation of the double series \re{A911} to
 \re{Kcirc} when inserted in the traces \re{defCAbb} is still an
 empirical observation awaiting a rigorous proof.}

 Our weak-coupling determinant formula, eqs.
 \re{octagonfinalpR0}-\re{seriesR}, provides an analytic expression
 for { the coefficients in the complete perturbative expansion of the
 octagon \re{defInOa}}.  The perturbative expansion of the octagon is
 expressed in terms of ladder integrals and at each order satisfies
 the maximal transcendentality requirement.

The expansion of the octagon in ladders can be arranged, as pointed
out by Coronado \cite{Coronado:2018cxj}, as a linear combination of
minors of the matrix of ladders \re{detf}.  The set of linearly
independent minors chosen in \cite{Coronado:2018cxj} and called there
Steinmann basis is made of minors having subsequent indices in the
vertical direction.  The expansion of the octagon in minors is a
direct consequence of our determinant representation
\re{octagonfinalpR}-\re{seriesR} when written in the form
\re{minoresR}.

 We hope that the determinant formula \re{DetOctbis} could give for
 the first time analytic access to the correlation functions at finite
 $g$.  It is likely that for the generic choice of the parameters one
 can truncate the matrix $\C\KKK$ to an $N\times N$ matrix with
 exponentially small in $N$ accuracy.  The CFT representation {of
 section 5 } in terms of free bosons of fermions is potentially useful
 in exploring the strong coupling regime, along the lines of
 \cite{JKKS2}.  The strong coupling expression should be given by the
 functional determinant of the quantum spectral curve in the sense
 discussed in \cite{JKKS2}.
 
 \bigskip
 \noindent
  {\bf Acknowledgements.}

\noindent The authors are obliged to Till Bargheer, Frank Coronado and
Pedro Vieira for discussions and useful exchanges.  This research is
partially supported by the Bulgarian NSF grant DN 18/1 and by the
bilateral grant STC/Bulgaria-France 01/6, PHC RILA 2018 N$^\circ$
38658NG.

\appendix

\section{ Perturbative evaluation of the matrix elements}
\label{AppendixA}

Here we derive formula \re{A911}.  The summation over $a$ and the
Fourier transformation in section \ref{sec:integralrep} can be done in
the reverse order leading to the same result as in \re{integralforK}
 \be
\la{integralforP} 
\begin{aligned}
&\CP_{m',m}
=i^{m-m'-1}  g \sum_a |z|^{1-a} {z^a-{\bar z}^a\over z-\bar z} 
\int _{ |\xi|}^\infty d t\   e^{-a t}\, 
  \({ \t_- \over \t_+}\)^{ m-m' \over 2}
J_{m+\ell}(2g\sqrt{\t_-\t_+})
 J_{m'+\ell}(2g\sqrt{\t_-\t_+}) \\
&
 ={ g\over 2 \, i} \int _{ |\xi|}^\infty d t\   \, 
 { \(i{ \sqrt{\t_- \over \t_+}}\)^{ m-m' }
J_{m+\ell}(2g\sqrt{\t_-\t_+})\, 
 J_{m'+\ell}(2g\sqrt{\t_-\t_+}) \over  \cosh t- \cos \phi}\,.
  \end{aligned} 
  \ee
  We next use that the 
 product of Bessel functions with the same argument
   can be 
 expanded in a power series, (8.442) in  \cite{GR}
 \be
 \la{productJJA}
 \begin{aligned}
J_m(2 y) J_{m'}(2 y)&
&=\sum_{p=0}^\infty { 2p +m+m' \choose p } { (-1)^p \, y^{m+m'+ 2p}
\over (m+p)!(m'+p)!}\,.
\end{aligned}
\ee
We insert this expansion for $y=g\, \sqrt{ \tau_+\, \tau_-}
 $ in the first line of \re{integralforP}, change variables $\tau_+
 \to t, \t_- \to t + 2|\xi|\, $ (assuming that $\xi=-\log \sqrt{z\bar
 z} >0$), expand $(t+ |2\xi|)^{\ell+m+p}$ and take the integral over
 $t$, (3.351-3.)  \cite{GR} \be
\begin{aligned}
\int_{|\xi |}^{\infty} e^{-at} \t_+^{\ell+m'+p}& \t_-^{\ell+m+p} =
(\ell+m'+p)!(\ell+m+p)!\times \\
 &    \sum_s 
{|\log z\bar z|^s\over s! } 
 {2\ell+m+m'+2p-s\choose \ell+m'+p} { |z|^a \over a^{2l+m+m'+2p-s +1}}  .
\end{aligned}
\ee
 The summation over $a$  in \re{integralforP} produces  
polylogarithms  
 \be
\la{defA4}
\begin{aligned}
\CP_{m',m}= |z| &i^{m-m'-1} \, \sum_{p= 0}^\infty (-1)^p { 2p + 2\ell+
m+m' \choose p } g^{ 2p+2\ell+ m+m'+1} \times \\ & \sum_{s=0}^{
p+m'+\ell} { 2 p+2\ell+m+m' -s \choose p+m'+\ell} \, { |\log z\bar
z|^s \over s!}\, \mathrm{L}_{ 2p+2\ell+ m+m'-s +1} \, \\
 \quad \text{with}& \quad \mathrm{L}_n= \mathrm{L}_n(z,\bar z) =
 {\mathrm{Li}_{n}(z) - \mathrm{Li}_{n}(\bar z)\over z-\bar z}.
 \end{aligned}
\ee
%
The integral in  
 \re{defCAshifted} for arbitrary $n$ after summation in $a_i, \varepsilon_i$
 reads (summation over all repeated indices assumed)

 \be {\rm Tr}(\C \KKK)^n= \prod_{i =1}^n C_{k_i m_i}\ (\CP_{m_{i-1}
 , k_i}-\CP_{k_i, m_{i-1}})\ , \, \ \ m_0=m_n  \,.\ee
For $n=1$ we get 
\be 
\la{defn1K}
\begin{aligned}
&-{1\over 2} {\rm Tr}(\C \KKK)=-\sum_{m\ge 0}\K_{m+1,m}= \ -\sum_{m\ge
0} (\CP_{m+1,m}-\CP_{m,m+1}) \\
&=e^{-\xi} \sum_{p\ge 0} \sum_{j\ge p+\ell+1} (-1)^{p} {2j-1\choose p}
g^{2j} \, F_j =e^{-\xi} \sum_{j\ge \ell+1} (-1)^{j-\ell-1}
{2j-2\choose j-\ell-1} g^{2j} \, F_j
\end{aligned}
\ee where $F_j$ is defined in \re{defF}.  The final expression in
\re{defn1K} reproduces $e^{-\xi}\, I_{1,\ell}$ in formula (C.1) of
\cite{Coronado:2018ypq}.

  Next we work out starting from \re{integralforP} the difference
  $K_{m, m+r}$ in \re{Kmndef}, for any $r\ge 1$.  For that one derives
  the identity \be\la{difid}
\begin{aligned}
&{1\over s!}\Big( {2a +r-s\choose a+r} +(-1)^{r -1}{2a +r-s\choose a}
\Big) \\
&=(-1)^{r-1}\sum_{k=1}^{[{r+1\over 2}]} {r-k\choose k-1}
{(a+k-1)!(a+k)!\over (s-r+2k-1)!  a! (a+r)!} {2a +2k - (s-
r+2k-1)\choose a+k -(s-r+2k+1)}
\end{aligned}
\ee applied for $a=p+m+\ell$.  Then changing the variable $s\to
s+r-2k+1$ the sum over $s$ in \re{defA4} reproduces the ladder
integral combination \re{defF} and we obtain \re{A911}.
 \footnotesize
 
%

\providecommand{\href}[2]{#2}\begingroup\raggedright\endgroup

\end{document}